\documentclass[aps,superscriptaddress,pra]{revtex4}
\usepackage{amsmath}
\usepackage{color}
\usepackage{graphicx}
\usepackage{amssymb}
\begin{document}

\title{Matter Wave Scattering and Guiding by Atomic Arrays}

\author{J. Y. Vaishnav}

\email{vaishnav@physics.harvard.edu}

\affiliation{Harvard University Department of Physics, Cambridge Massachusetts
02138}

\author{J. D. Walls}

\affiliation{Harvard University Department of Chemistry and Chemical Biology,
Cambridge Massachusetts 02138}

\author{M. Apratim}

\affiliation{Department of Electrical Engineering, Indian Institute of Technology,
Kharagpur 721302, India}

\author{E. J. Heller}

\affiliation{Harvard University Department of Physics, Cambridge Massachusetts
02138}

\affiliation{Harvard University Department of Chemistry and Chemical Biology,
Cambridge Massachusetts 02138}

\maketitle We investigate the possibility that linear arrays of atoms can guide matter waves, much as fiber optics guide light.  We model the atomic line as a quasi-1D array of $s$ wave point scatterers embedded in 2D.  Our theoretical study reveals how matter wave guiding arises from the interplay of scattering phenomena with bands and conduction along the array. We discuss the conditions under which a straight or curved array of atoms can guide a beam focused at one end of the array.

\section{Introduction\label{sec:Introduction}}

In this paper, we discuss the possibility of using a line of atoms
to guide matter waves, such as electrons or other atoms. Periodic
arrays often behave as waveguides; perhaps the most familiar example
is electrons propagating in a metal. Such waveguides can be
engineered as well, such as the guiding of electromagnetic waves in
a photonic crystal comprised of aluminum rods (see e.g.
\cite{JoannopoulosGuiding}). Although the problem of scattering from
periodic arrays is an old one, it arises now in a completely new
context, as recent technology allows structures to be engineered from
individual atoms. Single chains of Au atoms, for example, have
recently been deposited on NiAl(110) \cite{goldchains1,goldchains2}
as well as on Si(553) \cite{goldchains3}. Although our formalism
does not treat this situation specifically, artificial arrays of
atoms also arise when atoms are confined to individual sites in an
optical lattices. One might also imagine the individual scatterers
being an array of quantum dots, which could be used to guide
electrons.

We model the atomic  array as a set of $s$ wave scatterers. The
question of relevance from an experimental point of view is whether
we can use such an atomic array to guide waves. Theoretically, this
question translates into whether eigenstates of the system exist
which are propagating along but are evanescent transversely to the
array. In order to search for these states, we employ scattering
theory. Along the array, the system's periodicity gives rise to
Bloch waves, band structure, diffraction, and other features
familiar from solid state systems. However, as the array is finite
in the transverse direction, it acts like a partially transparent
wall in an unusual interplay between Bloch waves and scattering
theory: transmission and reflection coefficients, along with
resonant phenomena for propagation through the wall coexist with
conduction along the quasi-1D array. Although the same coexistence
exists for any real finite sample of periodic material, it is
especially evident and exposed in the quasi-1D periodic array.

In order for quasi-1D guiding to exist, the individual scatterers
making up the wall must be attractive. However, assuming elastic
scattering, a freely propagating incident mode can be captured and
guided along the array only if the array supports conducting states
at $E>0$.    It is easy to motivate the fact that that an array of
atoms (with each atom supporting an $E<0$ bound state) can  give
rise to an overall $E>0$ conducting state.

\begin{figure}
\begin{centering}\includegraphics[width=1\textwidth,keepaspectratio]{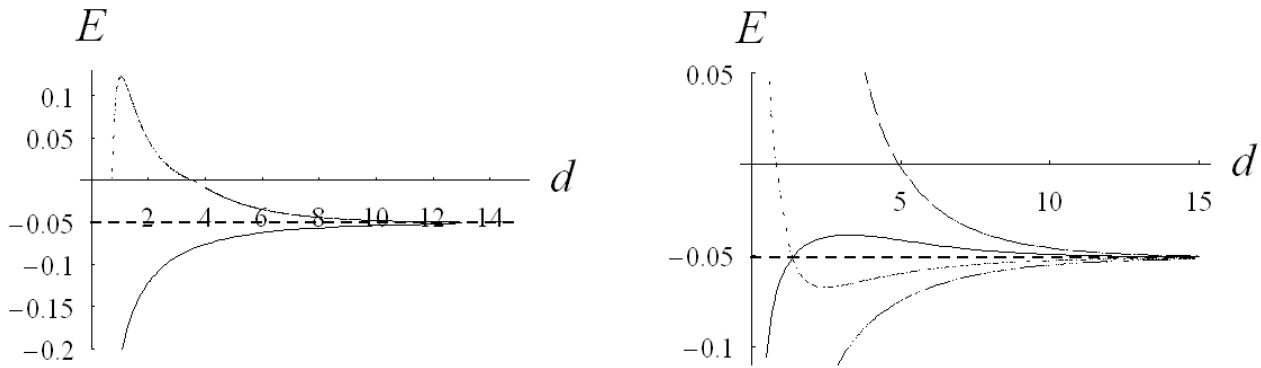}\par\end{centering}

\caption{Bound state energies for two scatterers (left) and four scatterers
(right) separated by distances $d$. The dashed line is the bound
state energy of an individual scatterer. For large $d$, the ground state is
the completely symmetric eigenstate, while the highest energy state is the completely
antisymmetric eigenstate. As the scatterers come closer together,
the antisymmetric eigenstate is pushed to positive energy. The scatterers
are point scatterers modelling the $s$ wave scattering from a cylindrical
well of depth $V_{0}=-0.8$ and radius $a=0.2$. \label{fig:threshold}}
\end{figure}

Consider a double well potential,
such that each individual well, when taken alone, admits a single
bound state. At large separation, the wells will give rise to a
degenerate doublet of symmetric and antisymmetric bound states. As
the wells come closer together, the antisymmetric state rises in
energy until it is pushed to $E>0$ and becomes a $p$ wave resonance,
as described in Ref. \cite{jessethreshold}, which is closely related
to the concept of proximity resonances \cite{HellerProxRes}. In Fig.
\ref{fig:threshold}, the bound state energies for two and four
scatterers are shown; as the scatterers come close together the
highest energy bound state is pushed
above threshold. %
For the infinite wall, one way to understand the presence of $E>0$
guided states for the wall is to consider the limit where the
scatterers are infinitely close together, thereby effectively forming
a potential trough of some width $w$:\[ V(x,y)=\left\{
\begin{array}{ccc}
-V_{0} &  & \left|x\right|<w\\
0 &  & \left|x\right|>w.\end{array}\right.\] Since the above
potential is separable, if the effective 1D well (about the
$\widehat{x}$ direction) admits an $E<0$ bound state, the trough
admits a continuum of $E>0$ states which are bound along the
$\widehat{x}$ direction, because any wavenumber can be taken in the
free $\widehat{y}$ direction. Thus by analogy to other, more
familiar systems, it seems reasonable that a wall of attractive
scatterers could behave as a waveguide.

We further note that at low energies, a curved chain of discrete, attractive atoms approximates a continuous, curved waveguide, suggesting an efficient numerical method for modeling continuous waveguides via multiple scattering methods.  A related method is the boundary wall method in \cite{boundarywall}.

The remainder of this paper is divided into five parts. We begin, in
Section \ref{sec:Foldy}, with a brief review of Foldy's method of
multiple scattering. In Section \ref{sec:Bloch}, we apply Foldy's
method to solve the Lippmann-Schwinger equation for a general
periodic array of scatterers. In Section
\ref{sec:Diffraction-and-Quasibound}, we investigate diffraction and
threshold resonances, uncovering a type of quasibound states which
are related to guiding. In Section \ref{sub:Bound-States} we
search for true guided states, and we calculate the band structure
for a single infinite wall of attractive scatterers. We demonstrate
that when individual scatterers have $E<0$ bound states, an array
can support conducting states at positive energies, $E>0.$ Finally,
in Section \ref{sec:Waveguides}, we numerically demonstrate that if
the array's symmetry is somehow broken, these conducting states
can be used to guide a beam; i.e., the array can capture an incident wave
focused on one end of the array, forcing it to propagate along the array and emerge on the other end.

\section{Background: Foldy's Method\label{sec:Foldy}}

The physical underpinning of all the resonance and interference
phenomena which we discuss in this paper is multiple scattering,
both within each unit cell and between different unit cells. To
model multiple scattering, we make extensive use of Foldy's method
\cite{Foldy}, which we briefly review here. Consider a wave
$\phi(\vec{r})$ incident on a collection of $N$ identical point
scatterers at positions $\left\{
\vec{r}_{1},\dots,\vec{r}_{N}\right\} $, where
$\vec{r}_{n}=(x_{n},y_{n})$. Applying the $t$ matrix formalism
\cite{rodbergthaler}, we characterize a single scatterer at
$\vec{r}_{i}$ by its $t$ matrix,\[
t=s(k)\left|\vec{r}_{i}\rangle\langle\vec{r}_{i}\right|\]
 where $s(k)$ is a function of the wavenumber, $k=\sqrt{2mE}/\hbar.$
The functional form of $s(k)$ is chosen to simulate the scatterer of
interest under the constraint that $s(k)$ must satisfy the optical
theorem

\begin{equation}
-\frac{2\hbar}{m}^{2}\mbox{Im}s(k)=\left|s(k)\right|^{2}.\label{eq:freeopticalfoldy}\end{equation}
 In the $t$ matrix formalism, the Lippmann-Schwinger equation for
multiple scattering is\begin{equation}
\psi(\vec{r})=\phi(\vec{r})+s(k)\sum_{i=1}^{N}\psi_{i}(\vec{r}_{i})G_{0}(\vec{r},\vec{r}_{i})\label{eq:lippmannSchwingerMultiple1}\end{equation}
 where
 \begin{equation}
G_{0}(\vec{r},\vec{r}_{0})=\frac{2m}{\hbar^{2}}\left[-\frac{i}{4}H^{(1)}_{0}(k\left|\vec{r}-\vec{r}_{0}\right|)\right]\label{eq:freegreens}
\end{equation}
 is the 2D free-space retarded Green's function satisfying\[
(\vec{\nabla}^{2}+k^{2})G_{0}(\vec{r},\vec{r}_{0})=\frac{2m}{\hbar^{2}}\delta^{(2)}(\vec{r},\vec{r}_{0}),\]
 and the various $\psi_{i}(\vec{r}_{i})$ in Eq. (\ref{eq:lippmannSchwingerMultiple1}) are defined recursively as\begin{equation}
\psi_{i}(\vec{r}_{i})=\phi(\vec{r}_{i})+s(k)\mathop{\sum_{j=1}^{N}}_{j\neq i}
\psi_{j}(\vec{r}_{j})G_{0}(\vec{r}_{i},\vec{r}_{j}).\label{eq:lippmanSchwingerMultiple2}\end{equation}
$\psi_{i}(\vec{r}_{i})$ is the \emph{effective} incoming
wavefunction evaluated at the $i^{{\textrm{th}}}$ scatterer after
scattering from each of the other scatterers, excluding the singular
self-interaction of the $i^{{\textrm{th}}}$ scatterer. In the
remainder of this paper, we set $\hbar=m=1$ and denote $s(k)$ as $s$.

Defining two $N$ x 1 column vectors, $\vec{\phi}$ and $\vec{\psi}$
,whose $i^{{\textrm{th}}}$ elements are given by
$\vec{\phi}_{i}\equiv\phi(\vec{r}_{i}),$
$\vec{\psi}_{i}\equiv\psi(\vec{r}_{i}),$ the Lippmann-Schwinger
equation can be written as a simple matrix equation by inverting
Eqs.
(\ref{eq:lippmannSchwingerMultiple1})-(\ref{eq:lippmanSchwingerMultiple2})
to yield\begin{equation}
\vec{\psi}=\mathbf{M}^{-1}\vec{\phi}\label{eq:lippmanSchwingerMatrix}\end{equation}
 where \[
\mathbf{M}\equiv1-s\mathbf{G}\]
 and the matrix $\mathbf{G}$ is defined by\begin{equation}
G_{ij}\equiv\left\{ \begin{array}{cc}
0 & i=j\\
G_{0}(\vec{r_{i}},\vec{r}_{j}) & i\neq j.\end{array}\right.\label{eq:gmatrix}\end{equation}
 $\mathbf{G}$ excludes the singular self-interactions of each scatterer
with itself.

Substituting the values of $\vec{\psi}_{i}$ from Eq. (\ref{eq:lippmanSchwingerMatrix})
into Eq. (\ref{eq:lippmannSchwingerMultiple1}) yields an expression
for the scattered wavefunction:\begin{equation}
\psi(\vec{r})=\phi(\vec{r})+s\sum_{i=1}^{N}G_{0}(\vec{r},\vec{r}_{i})\left(\mathbf{M}^{-1}\vec{\phi}\right)_{i}.\label{eq:finalfreems}\end{equation}

\section{Multiple Scattering From General Periodic Structures: Connection
to Bloch waves\label{sec:Bloch}}

The problem of guiding is related to finding conducting eigenstates
for an array of scatterers. We discussed in Section
\ref{sec:Introduction} how Bloch waves and scattering phenomena
coexist in such systems. The usual 3D approach of reciprocal lattices could be adapted to 2D. We instead apply multiple scattering theory, considering all of
the multiple scattering events within each unit cell as well as
between different unit cells. The multiple scattering approach,
although more involved, yields detailed information about
interference processes and also generalizes more easily to the
introduction of disorder into the lattice.

\begin{figure}
\begin{centering}\includegraphics[scale=0.5]{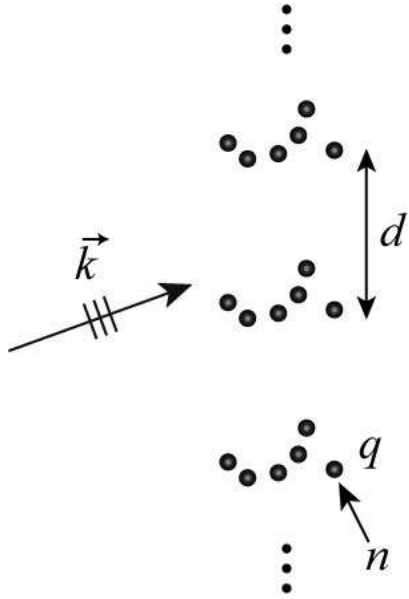}\par\end{centering}

\caption{Plane wave incident on array with Bravais
vector $d\hat{y}.$  The unit cell is indexed by $q$, and each
individual scatterer is indexed by $n.$ \label{fig:lattice}}
\end{figure}

In this section, we apply Foldy's method to solve the
Lippmann-Schwinger equation for a plane wave scattering from a
general, infinite periodic array of clusters of point scatterers
with a Bravais lattice spanned by $d\hat{y}$ (Fig.
\ref{fig:lattice}).  The multiple scattering solution in Eq.
(\ref{eq:lippmanSchwingerMatrix}) would seem to require inversion of
a biinfinite matrix. In this section, however, we reduce the
solution to inversion of an $N\times N$ matrix, where $N$ is the
number of scatterers per unit cell, and demonstrate how Bloch waves
arise naturally from multiple scattering theory. The resulting
Lippmann-Schwinger equation resembles Eqs.
(\ref{eq:lippmanSchwingerMatrix})-(\ref{eq:gmatrix}) but with an
effective scattering strength $\tilde{s}$ and an effective Green's
matrix $\tilde{\mathbf{G}}$ which account for multiple scattering
between unit cells.

Our approach turns out to be related to the Korringa-Kohn-Rostoker
(KKR) method \cite{KKR1,KKR2}, with the variation that we have
represented our scatterers by $t$ matrices and have begun with an
$s$ wave approximation. A related approach has been applied to study
scattering from two-dimensional periodic slabs of scatterers
embedded in three dimensions \cite{arraymultiscatter}.

\subsection{Multiple Scattering from a Periodic Grating}

Denote by $\vec{r}_{n}^{(q)}$ the position of the
$n^{{\textrm{th}}}$ of $N$ scatterers in unit cell $q.$   Further
suppose that all $N$ scatterers in a unit cell are identical with
$t$ matrix given by $s$ (the arguments in this section can easily be
generalized to the case of nonidentical scatterers). Applying
Foldy's method, we can write the Lippmann-Schwinger equation
as\begin{eqnarray}
\psi(\vec{r}) & = & \phi(\vec{r})+s\sum_{q=-\infty}^{\infty}\sum_{n=1}^{N}G_{0}(\vec{r},\vec{r}_{n}^{(q)})\psi_{n}^{(q)}(\vec{r}_{n}^{(q)})\label{eq:foldyarray1}\\
\psi_{n}^{(q)}(\vec{r}_{n}^{(q)}) & = &
\phi(\vec{r}_{n}^{(q)})+s\sum_{p=-\infty}^{\infty}
\mathop{\sum_{m=1}^N}_{(m,p)\neq (n,q)}
G_{0}(\vec{r}_{m}^{(p)},\vec{r}_{n}^{(q)})\psi_{m}^{(p)}(\vec{r}_{m}^{(p)})\label{eq:foldyarray2}\end{eqnarray}
 where $G_{0}(kr)$ is the free space Green's function ($\hbar=m=1)$ from
Eq. (\ref{eq:freegreens}). Once again, if the effective incident
wavefunction amplitudes at the scatterers,
$\psi_{n}^{(q)}\left(\vec{r}_{n}^{(q)}\right)$ in Eq.
(\ref{eq:foldyarray2}), are known, the full wave function is
determined using Eq. (\ref{eq:foldyarray1}).

It is straightforward to recursively sum and reindex Eq.
(\ref{eq:foldyarray2}) to show that for an incident wave (normalized
to unit flux along the $\widehat{x}$ direction):
\[
\phi(\vec{r})=\frac{1}{\sqrt{k^{(0)}_{x}}}e^{i\vec{k}\cdot\vec{r}}\]
 the solutions to Eq. (\ref{eq:foldyarray2}) are reduced to finding
 the wave function amplitudes for a single unit cell, since\begin{equation}
\psi_{n}^{(q)}(\vec{r}_{n}^{(q)})=e^{ik_{y}qd}\psi_{n}^{(0)}(\vec{r}_{n}^{(0)}).\label{eq:blochrelation}\end{equation}
Let us focus on the $N$ scatterers in the unit cell indexed by
$q=0.$ Using Eq. (\ref{eq:blochrelation}) on the right hand side of
Eq. (\ref{eq:foldyarray2}), evaluated for $q=0,$ we
obtain\begin{equation}
\psi_{n}^{(0)}(\vec{r}_{n}^{(0)})=\phi(\vec{r}_{n}^{(0)})+s \sum_{p=-\infty}^{\infty}
\mathop{\sum_{m=1}^{N}}_{(m,p)\neq(n,0)}
G_{0}(\vec{r}_{m}^{(p)},\vec{r}_{n}^{(0)})e^{ikypd}\psi_{m}^{(0)}(\vec{r}_{m}^{(0)})\label{eq:blochrelation2}\end{equation}
We wish to separate off the $m=n$ term, which corresponds to
multiple scattering between each scatterer and its periodic
counterparts in other unit cells. It is in this term that we must
exclude the self interaction, which corresponds to $m=n$ and $p=0$.
Breaking up the sum, we find \begin{eqnarray}
\psi_{n}^{(0)}(\vec{r}_{n}^{(0)}) & = &
\phi(\vec{r}_{n}^{(0)})+s\mathop{\sum_{p=-\infty}^{\infty}}_{p\neq0}G_{0}(k\left|p\right|d)e^{ikypd}\psi_{n}^{(0)}(\vec{r}_{n}^{(0)})\nonumber\\
 &  & +s\mathop{\sum_{m=1}^{N}}_{m\neq
 n}\left[\sum_{p=-\infty}^{\infty}G_{0}(\vec{r}_{m}^{(0)}+pd\hat{y},\vec{r}_{n}^{(0)})e^{ikypd}\right]\psi_{m}^{(0)}(\vec{r}_{m}^{(0)})\nonumber\\
 & = & \phi(\vec{r}_{n}^{(0)})+sG_{r}\psi_{n}^{(0)}(\vec{r}_{n}^{(0)})+s\sum_{m=1}^{N}\mathbf{\tilde{G}}_{mn}\psi_{m}^{(0)}(\vec{r}_{m}^{(0)})\label{eq:grenorm}\end{eqnarray}
In Eq. (\ref{eq:grenorm}), $G_{r}$ is a scalar quantity independent
of the configuration of the unit cell and is given
by\begin{equation} G_{r}\equiv
s\mathop{\sum}_{p\neq0}G_{0}(k\left|p\right|d)e^{ik_{y}pd}\label{eq:renormalized}\end{equation}
 and $\mathbf{\tilde{G}}$ is an $N\times N$ matrix defined as\begin{equation}
\mathbf{\tilde{G}}_{mn}=\left\{ \begin{array}{ccc}
0 &  & m=n\\
G(\vec{r}_{n}^{(0)}-\vec{r}_{m}^{(0)}) &  & m\neq
n\end{array}\right.\label{eq:grenormalized}\end{equation} where we
used the lattice sum\begin{equation}
G(\vec{r})=\sum_{p=-\infty}^{\infty}G_{0}(\vec{r},pd\hat{y})e^{ik_{y}pd}\label{eq:latticehankelsum}\end{equation}
A rapidly converging expression for $G(\vec{r})$ can be found in Eq.
(\ref{eq:planegreens}) of Appendix \ref{sec:Relevant-Lattice-Sums}.

Defining vectors of wavelets \begin{eqnarray} \vec{\phi}^{(0)} & = &
\left(\begin{array}{c}
\phi_{1}^{(0)}(\vec{r}_{1}^{(0)})\\
\phi_{2}^{(0)}(\vec{r}_{2}^{(0)})\\
\vdots\\
\phi_{N}^{(0)}(\vec{r}_{N}^{(0)})\end{array}\right),\nonumber\\
\vec{\psi}^{(0)} & = & \left(\begin{array}{c}
\psi_{1}^{(0)}(\vec{r}_{1}^{(0)})\\
\psi_{2}^{(0)}(\vec{r}_{2}^{(0)})\\
\vdots\\
\psi_{N}^{(0)}(\vec{r}_{N}^{(0)})\end{array}\right)\label{eq:wavelets1}\end{eqnarray}
 we can solve for the wavelets in the $q=0$ unit cell:\begin{eqnarray}
\vec{\psi}^{(0)} & = & \frac{1}{1-sG_{r}}\tilde{\mathbf{M}}^{-1}\vec{\phi}^{(0)}\label{eq:blochsum}\end{eqnarray}
 where we have defined\begin{equation}
\tilde{\mathbf{M}}=\mathbf{I}-\tilde{s}\tilde{\mathbf{G}.}\label{eq:mmatrix}\end{equation}

and

\begin{equation}
\tilde{s}=\frac{s}{1-sG_{r}}\label{eq:stwiddle}\end{equation}
 The remaining wavelets in other unit cells are then simply determined by Eq. (\ref{eq:blochrelation}):\begin{equation}
\vec{\psi}^{(q)}=e^{ik_{y}qd}\vec{\psi}^{(0)}.\label{eq:blochwavelets}\end{equation}
 The full wavefunction is finally given by substituting Eq. (\ref{eq:blochsum})
and Eq. (\ref{eq:blochwavelets}) into Eq.
(\ref{eq:foldyarray1}):\begin{eqnarray}
\psi(\vec{r}) & = & \phi(\vec{r})+\tilde{s}\sum_{p=-\infty}^{\infty}\sum_{n=1}^{N}G_{0}(\vec{r},\vec{r}_{n}^{(p)})e^{ik_{y}pd}\left(\tilde{\mathbf{M}}^{-1}\vec{\phi}^{(0)}\right)_{n}\label{eq:blochscatteredwave}\\
 & = & \phi(\vec{r})+\tilde{s}\sum_{n=1}^{N}\left(\tilde{\mathbf{M}}^{-1}\vec{\phi}^{(0)}\right)_{n}\sum_{p=-\infty}^{\infty}G_{0}(\vec{r}-\vec{r}_{n}^{(0)},pd\hat{y})e^{ik_{y}pd}\\
 & = & \phi(\vec{r})+\tilde{s}\sum_{n=1}^{N}G(\vec{r}-\vec{r}_{n}^{(0)})\left(\tilde{\mathbf{M}}^{-1}\vec{\phi}^{(0)}\right)_{n}\label{eq:blochscatteredwave3}\end{eqnarray}
 Note the similarity between Eq. (\ref{eq:blochscatteredwave3}) and
Eq. (\ref{eq:finalfreems}):  one can go from a single unit cell to a
repeating array simply by replacing the $t$ matrix $s(k)$ with its
renormalized version $\tilde{s}(\vec{k})$, and the free space
Green's function $G_{0}(\vec{r})$ with the effective Green's
function $G(\vec{r})$. The wavefunction in Eq.
(\ref{eq:blochscatteredwave}) is a Bloch wave since\[
\psi(\vec{r}+d\hat{y})=e^{ik_{y}d}\psi(\vec{r}).\] We note that the
renormalization and interference effects in a periodic grating are
very similar to the effects encountered when a scatterer or a
cluster of scatterers is placed in an external confining potential
(see e.g. \cite{MyPaper, Olshanii}).  The similarity arises because
scattering in a confined geometry is also effectively a multiple
scattering problem:   A particle can scatter once from the target,
reflect from the confining potential, and scatter again.  In the
case of a cluster of scatterers confined to a hard walled or
periodic waveguide, the mapping to an array is in fact explicit;
applying the method of images, the cluster becomes an infinite array
where the effective ``unit cells" are images of the confined cluster
of scatterers.

\section{\label{sec:Diffraction-and-Quasibound}Diffraction and Quasibound
States}

The result in Eq. (\ref{eq:blochscatteredwave3}) is in terms of
superpositions of spherical waves. Alternatively, Eq.
(\ref{eq:blochscatteredwave3}) could be written in a basis of plane
and evanescent waves. Substituting Eq. (\ref{eq:planegreens}) into
Eq. (\ref{eq:blochscatteredwave3}) yields a plane plus evanescent
wave expansion for the scattered wave:\begin{equation}
\psi(\vec{r})=\frac{e^{i\vec{k}_{0}\cdot\vec{r}}}{\sqrt{k_{x}^{(0)}}}-\frac{i\tilde{s}}{d}\sum_{q=-\infty}^{\infty}\sum_{n=1}^{N}\left(\mathbf{\tilde{M}}^{-1}\vec{\phi}^{(0)}\right)_{n}\frac{1}{k_{x}^{(q)}}e^{ik_{x}^{(q)}\left|x-x_{n}^{(0)}\right|}e^{ik_{y}^{(q)}(y-y_{n}^{(0)})}\label{eq:scatteredwavetr}\end{equation}
where the wavenumbers of the diffracted beams are quantized by the
Bragg condition:\begin{eqnarray}
k_{y}^{(q)} & = & k_{y}^{(0)}+\frac{2q\pi}{d}\label{eq:braggarray1}\\
k_{x}^{(q)} & = & \sqrt{k^{2}-\left(k_{y}^{(q)}\right)^{2}.}\label{eq:braggarray2}\end{eqnarray}
The values of $q$ for which $k_{x}^{(q)}$ is real correspond to
diffracted plane waves, while the remaining values of $q$ correspond
to evanescent waves. The set $\mathcal{Q}$ of open channels, corresponding
to diffracted plane waves, is defined by $\mathcal{Q}=\left[Q_{min},Q_{max}\right]$
where \begin{eqnarray}
Q_{min} & = & \left\lceil -\frac{(k+k_{y})d}{2\pi}\right\rceil \label{eq:nmin}\\
Q_{max} & = & \left\lfloor \frac{(k-k_{y})d}{2\pi}\right\rfloor
.\label{eq:nmax}\end{eqnarray} Eq. (\ref{eq:scatteredwavetr}) is in
essence Bragg diffraction: the scattered wave consists of a finite
number of diffracted plane waves propagating at the Bragg angles
given by Eqs. (\ref{eq:braggarray1})-(\ref{eq:braggarray2}) and an
infinite number of evanescent waves.

In the far field, defined by $\left|x\right|\gg\left|x_{n}\right|$
for all $n$, only the diffracted beams survive, and the transmitted
and reflected wavefunctions are\begin{eqnarray}
\psi_{R}(\vec{r}) & = & \sum_{q\in\mathcal{Q}}R_{q}\frac{1}{\sqrt{k_{x}^{(q)}}}e^{-ik_{x}^{(q)}x+ik_{y}^{(q)}y}\\
\psi_{T}(\vec{r}) & = &
\sum_{q\in\mathcal{Q}}T_{q}\frac{1}{\sqrt{k_{x}^{(q)}}}e^{i k_{x}^{(q)}x+ik_{y}^{(q)}y}\end{eqnarray}
 where the reflection and transmission coefficients, $R_{q}$ and $T_{q}$, are given by (using Eq. (\ref{eq:scatteredwavetr}))\begin{eqnarray}
R_{q} & = & -\frac{i\tilde{s}}{d}\times\frac{1}{\sqrt{k_{x}^{(q)}}}\sum_{n=1}^{N}(\tilde{\mathbf{M}}^{-1}\vec{\phi}^{(0)})_{n}e^{ik_{x}^{(q)}\left|x_{n}^{(0)}\right|}e^{-ik_{y}^{(q)}y_{n}^{(0)}}.\label{eq:rn}\\
T_{q} & = &
\delta_{q}-\frac{i\tilde{s}}{d}\times\frac{1}{\sqrt{k_{x}^{(q)}}}\sum_{n=1}^{N}(\tilde{\mathbf{M}}^{-1}\vec{\phi}^{(0)})_{n}e^{-ik_{x}^{(q)}\left|x_{n}^{(0)}\right|}e^{-ik_{y}^{(q)}y_{n}^{(0)}}.\label{eq:tn}\end{eqnarray}
The quantities $\left|R_{q}\right|^{2}$ and $\left|T_{q}\right|^{2}$
correspond to the probability of reflection or transmission into the
$q^{\mbox{th}}$ mode; the $q=0$ mode is the specular component. The
total reflection and transmission probabilities for the wall of
scatterers are given by\begin{eqnarray}
R & = & \sum_{q\in\mathcal{Q}}\left|R_{q}\right|^{2}\label{eq:refarray}\\
T & = &
\sum_{q\in\mathcal{Q}}\left|T_{q}\right|^{2}.\label{eq:transarray}\end{eqnarray}
The value of renormalized $t$ matrix, $\tilde{s}$, is constrained by
combining Eq. (\ref{eq:refarray}) and Eq. (\ref{eq:transarray}) with
the unitarity requirement $R+T=1.$ This constraint is an analog of
an optical theorem for the grating. The optical theorem for the
single wall of scatterers is derived in Appendix
\ref{sec:Unitarity-and-The}.

\subsection{Threshold Resonances and Quasibound States\label{sub:Mechanisms-for-Guiding:}}

\begin{figure}
\begin{centering}\includegraphics[width=1\textwidth,keepaspectratio]{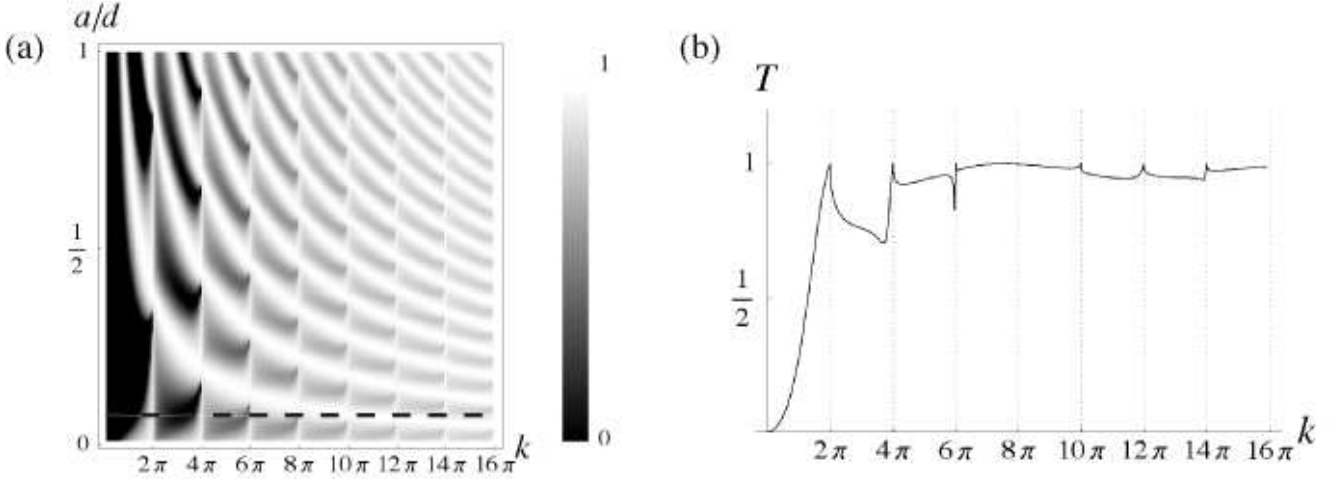}\par\end{centering}

\caption{(a) Transmission coefficient for scattering of an incident beam with
wavenumber $k$ from a wall of hard disks with radius $a.$ The incident
beam is at normal incidence $(k_{y}^{(0)}=0).$ At low energies, the
beam is fully reflected; at high energies the beam is almost entirely
transmitted. Note the transmission resonances at $k=k_{y}^{(n)}.$
(b) Structure of the resonances at $a=0.1$ (dashed line in (a)).
 This figure also illustrates the breakdown of the $s$ wave approximation
at high energies; classically, the limit of the transmission should
not be exactly 1.\label{fig:transmission}}
\end{figure}

Since the signature of a bound or quasibound state is often a
transmission resonance, we begin by examining the transmission
coefficient of the array, Eq. (\ref{eq:transarray}). As illustrated
in Fig. \ref{fig:transmission} for the single chain of atoms,
resonances occur at energies such that $k_{x}^{(q)}\rightarrow0$.
Such resonances, called threshold resonances, correspond to energies
at which one of the evanescent waves becomes a propagating
diffracted beam. These threshold resonances are purely due to the
periodicity of the array, and the resonance energies are independent
of the type or configuration of the scatterers within an individual
unit cell. Threshold resonances occur in atom-surface scattering as
selective adsorption resonances \cite{selectiveadsorption}, in X-ray
diffraction as emergent beam resonances, and in acoustics as Parker
resonances \cite{RayleighExpt}.

Near a threshold resonance, i.e., as $k\rightarrow
k_{y}^{(q)}\pm\epsilon$, the lattice sum in Eq. (\ref{eq:grwall})
yields
\begin{equation}
\lim_{k=k_{y}^{(q)}\pm\epsilon}G_{r}=-\frac{i}{d}\frac{1}{\sqrt{\pm2k_{y}^{(q)}\epsilon}}.\label{eq:limgrarray}\end{equation}
 From Eq. (\ref{eq:stwiddle}) and Eq. (\ref{eq:limgrarray}),
\begin{equation}
\lim_{k=k_{y}^{(q)}\pm\epsilon}\tilde{s}=\frac{1}{G_{r}}=id\sqrt{\pm2k_{y}^{(q)}\epsilon}.\label{eq:limitstwwall1}\end{equation}
From Eq. (\ref{eq:limitstwwall1}), the dependence of $\tilde{s}$ on
the bare $t$ matrix $s(k)$ cancels entirely near the threshold
resonance. Exactly on threshold, $\tilde{s}=0$, which from Eqs.
(\ref{eq:rn})-(\ref{eq:tn}) yields $T=1$;  the array becomes
entirely transparent, and the incident beam is entirely transmitted
(although with a phase). This transparency, which is a form of the
Ramsauer-Townsend effect, is consistent with flux conservation, as
the guided beam carries no flux away from the array.

\begin{figure}
\begin{centering}\includegraphics[width=0.7\textwidth,keepaspectratio]{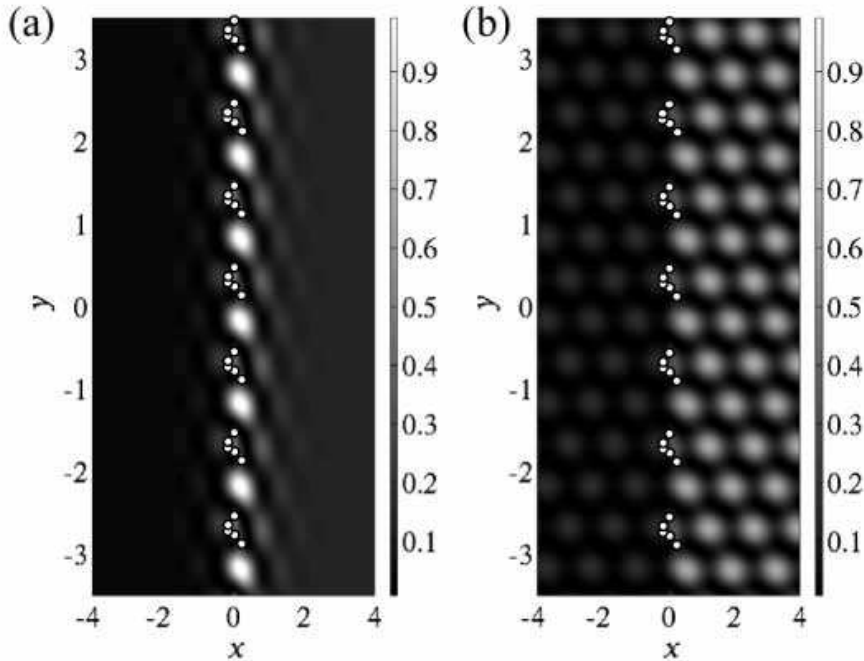}\par\end{centering}

\caption{Threshold resonance for a unit cell with
five randomly placed scatterers (hard disks with $a=0.1$) depicted by
the dots in the figure). The full wavefunction just above and
below threshold is given by Eqs. (\ref{eq:quasiarray1},
\ref{eq:quasiarray2}). For clarity, we plot only the scattered
wavefunction; the incident wave is a plane wave and would be added
to the scattered wavefunction to obtain the full wavefunction. (a)
The quasibound state just below resonance, at wavenumber
$kd=2\pi-0.3$, becomes unbound (b) just above resonance, at
$kd=2\pi+0.3$. \label{fig:arraytrapped}}
\end{figure}

Inserting Eq. (\ref{eq:limitstwwall1}) into Eq.
(\ref{eq:scatteredwavetr}), the scattered wave near threshold
consists entirely of the emergent beam traveling along the $\hat{y}$
axis: \begin{eqnarray}
\lim_{k=k_{y}^{(q)}-\epsilon}\psi(\vec{r}) & = & e^{i\vec{k}\cdot\vec{r}}+\sum_{n=1}^{N}\left(\tilde{\mathbf{M}}^{-1}\vec{\phi}^{(0)}\right)_{n}e^{-\sqrt{2k_{y}^{(q)}\epsilon}\left|x-x_{n}^{(0)}\right|}e^{ik_{y}^{(q)}(y-y_{n}^{(0)})}+{\textrm{O}}(\epsilon^{1/2})\label{eq:quasiarray1}\\
\lim_{k=k_{y}^{(q)}+\epsilon}\psi(\vec{r}) & = &
e^{i\vec{k}\cdot\vec{r}}+\sum_{n=1}^{N}\left(\tilde{\mathbf{M}}^{-1}\vec{\phi}^{(0)}\right)_{n}e^{i\sqrt{2k_{y}^{(q)}\epsilon}\left|x-x_{n}^{(0)}\right|}e^{ik_{y}^{(q)}(y-y_{n}^{(0)})}+{\textrm{O}}(\epsilon^{1/2})\label{eq:quasiarray2}\end{eqnarray}
Fig. \ref{fig:arraytrapped} illustrates the probability density
above (Fig. \ref{fig:arraytrapped}(b)) and below (Fig.
\ref{fig:arraytrapped}(a)) a threshold resonance for a wall of hard
disks. For clarity, the incident wavefunction
$e^{i\vec{k}\cdot\vec{r}}$ is omitted from $\psi(\vec{r})$ in Eqs.
(\ref{eq:quasiarray1})-(\ref{eq:quasiarray2}) when plotting
$|\psi(\vec{r})|^{2}$ in Fig. \ref{fig:arraytrapped}.  Approaching
threshold from below, the evanescent beam, which is about to emerge,
dominates the scattering. The scattered state consists almost
entirely of a state which is evanescent along the $\hat{x}$
direction (Fig. \ref{fig:arraytrapped}a) but becomes progressively
more weakly bound as we approach threshold. At threshold, the
scattered state merges with the continuum, and for wavenumbers just
above threshold, the scattered state is weakly unbound in the
$\hat{x}$ direction (Fig. \ref{fig:arraytrapped}b). The threshold
resonance thus corresponds to quasibound states which conduct along
the wire.  Although these quasibound states constitute, in some
sense, a form of guiding, they are not truly conducting states. The
$\mbox{O}(\epsilon^{1/2})$ term in Eqs.
(\ref{eq:quasiarray1})-(\ref{eq:quasiarray2}) is due to coupling to
other unbound states. The $\psi(\vec{r})$ states in Eqs.
(\ref{eq:quasiarray1})-(\ref{eq:quasiarray2}) are thus quasibound
rather than being truly bound, and they have finite lifetimes.
 Similar states, known as Rayleigh-Bloch waves, exist in many other
physical systems ranging from ocean coastlines \cite{RayleighCoast}
to acoustics \cite{RayleighExpt}.

\section{\label{sub:Bound-States}Conducting States}

Motivated by the presence of quasibound states, we now refine our
search to find genuine conducting eigenstates. An array of
attractive scatterers, in the limit where the scatterers are closely
spaced, should resemble a quasi-1D potential trough embedded in 2D
and give rise to a set of states which are purely bound along the
array. Such conducting states would be evanescent in the $\hat{x}$
direction but free in the $\hat{y}$ direction and thus correspond to
states with wavevectors\begin{equation}
\vec{k}=i\kappa_{x}\hat{x}+k_{y}\hat{y}\label{eq:boundwavev}\end{equation}
In contrast to the quasibound states in Section
\ref{sub:Mechanisms-for-Guiding:}, the truly conducting states would
contain \emph{only} such wavevectors; in a time dependent sense;  a
wavepacket injected into such a state would conduct forever. Unlike
the quasibound states, bound states depend strongly on the
properties of the grating and do not necessarily exist for an
arbitrary periodic grating.  For example, a periodic array of
repulsive scatterers, while possessing infinitely many quasibound
states, would
not have any truly conducting states. %

For a single point scatterer in free space, two criteria
characterize a bound state: the state is (1) localized (has negative
energy) and (2) is a pole of the $t$ matrix (the interpretation of
this criterion is that the bound state is a scattering state which
exists even in the absence of an incoming wavefunction).  For an
array of scatterers, Foldy's method transforms the
Lippmann-Schwinger equation into the matrix equation, Eq.
(\ref{eq:blochsum}), which can be rewritten as\begin{eqnarray}
\left[\left(1-sG_{r}\right)\mathbf{\tilde{M}}\right]^{-1}\vec{\psi}^{(0)}
& = & \vec{\phi}^{(0)}.\label{eq:blochsum2}\end{eqnarray} The
existence of a scattered wavefunction for zero incoming wavefunction
implies the existence of a homogeneous solution to Eq.
(\ref{eq:blochsum2}). States which exist in the absence of an
incoming wave must exist at values of $\vec{k}$ which are roots of
the secular equation \begin{equation}
(1-sG_{r})^{N}\det\mathbf{\tilde{M}}=0.\label{eq:secular}\end{equation}
This secular equation is strongly dependent on the configuration of
the unit cell and must be solved for a particular grating. We shall
here focus on the single wall of atoms, which is the most common
experimental setup \cite{goldchains1,goldchains2,goldchains3}. For
the single wall, $\tilde{\mathbf{M}}=\mathbf{I},$ and Eq.
(\ref{eq:secular}) simplifies to\begin{equation}
1-sG_{r}=0,\label{eq:boundcriterion1}\end{equation} implying, from
Eq. (\ref{eq:stwiddle}), that a bound state corresponds to a pole of
the renormalized $t$ matrix, $\tilde{s}$.

\begin{figure}
\begin{centering}\includegraphics[width=1\textwidth,keepaspectratio]{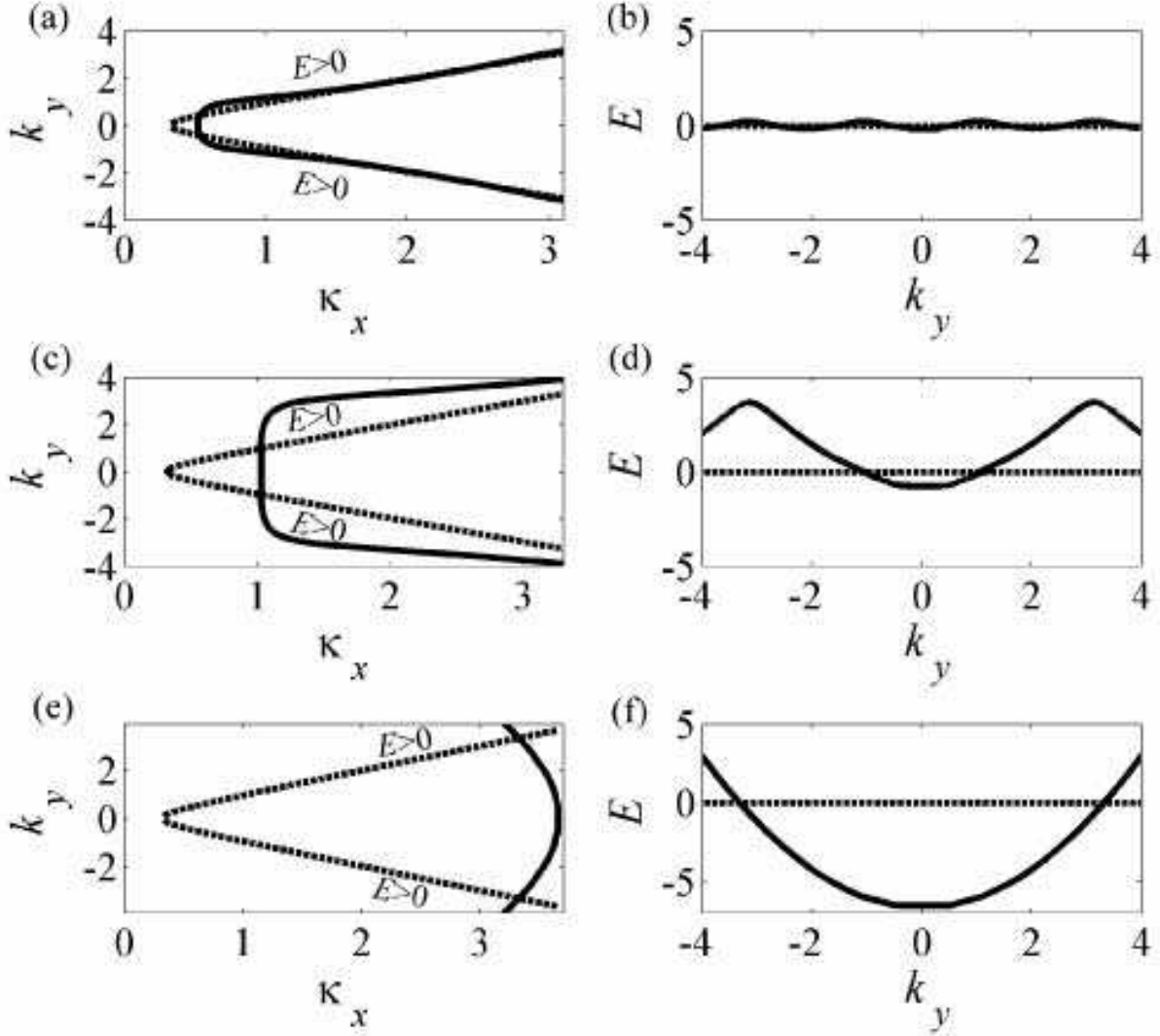}\par\end{centering}

\caption{(a,c,e) show the curves along which $\tilde{s}(i\kappa_{x},k_{y})$
has poles (the symmetry across $k_{y}=0$ is due to reflection symmetry
of the array along $y=0$). These curves correspond to manifolds of
conducting states bound in the $x$ direction, and propagating along the array.
The dashed curves, for comparison, are contours of wavevectors corresponding
to bound state energy of a single scatterer. (b,d, f) show the band
structure ($E$ vs. $k_{y}$) for the conducting states. (a,b) correspond
to a lattice constant $d=3.0$; $(c,d)$ to $d=$1.0, and (e,f) to
$d=0.202$. As the wells approach each other, the bands are increasingly
perturbed from the single scatterer limit. In all cases, conducting states
are present at $E>0$. \label{fig:wallpole}}

\end{figure}

\begin{figure}
\begin{centering}\includegraphics[width=0.4\textwidth]{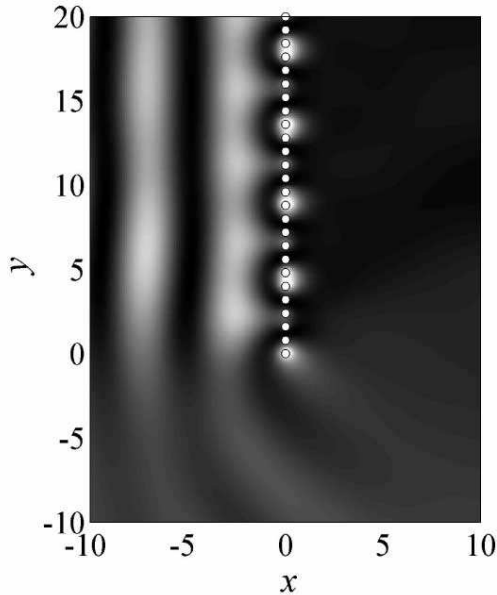}\par\end{centering}

\caption{Plane wave at normal incidence ($k=0.7)$
scattering into a conducting state; scatterers (depicted by dots
in the figure) are soft disks with $V_{0}=-8.0$, $a=0.2$ and
$d=0.8$ (the same scatterers as in Figs. \ref{fig:threshold} and
\ref{fig:wallpole}). This figure is a numerical demonstration that a
propagating incident wave can couple into an $E>0$ conducting state
of the array, if the translational invariance of the array is
broken. Coupling into the conducting state with a plane wave does
not constitute guiding, as the incident wave already has amplitude
everywhere.\label{fig:threshie}}
\end{figure}

In order to find conducting states, we evaluate $\tilde{s}$ on a
grid of values of $(\kappa_{x},k_{y})$, and numerically search for
poles of the form given in Eq. (\ref{eq:boundwavev}) with
corresponding energy\[
E=\frac{1}{2}\sqrt{-\kappa_{x}^{2}+k_{y}^{2}}.\] Bound states with
$E>0$ correspond to those poles of the renormalized $t$ matrix for
which $\left|k_{y}\right|>\left|\kappa_{x}\right|.$  We used the $t$
matrix in Eq. (\ref{eq:stangent}) to simulate an infinite wall of
attractive cylindrical wells with depth $V_{0}=-8.0$ and radius
$a=0.2,$ so that an individual well allowed a single bound state
just barely below threshold (at $E=-0.05$, or
$\kappa=-\sqrt{2E}=0.32i$). Fig. \ref{fig:threshold} depicts how
arrays of two or four such wells support bound states above $E=0$ as
the wells are placed close together.  Figs.
\ref{fig:wallpole}(a,c,e) show the manifold of poles in $\vec{k}$
space for the infinite wall of scatterers for three different
lattice constants. In each figure, the solid curves correspond to
the manifolds of conducting states for the array, and the dashed
curves correspond to the bound state energies of a single scatterer,
given by \[ -\kappa_{x}^{2}+k_{y}^{2}=-\kappa^{2}.\] A more
traditional view of the band structure for each potential is
presented in Figs. \ref{fig:wallpole}(b, d, f), where the energies
of the conducting states versus $k_{y}$ have been plotted. Figs.
\ref{fig:wallpole}(a,b) correspond to a large well separation
($d=3.0)$, Figs. \ref{fig:wallpole}(c,d) to an intermediate well
separation $(d=1.0)$, and Figs. \ref{fig:wallpole}(e,f) to a small
well separation $(d=0.202).$ When the wells are far apart, the
collective bound state is near the bound state energy of a single
scatterer, but as the wells approach each other, bound states
occurring at $E>0$ start to appear.  These states are particularly
interesting since they could, in principle, be accessed by a
propagating incident wave.  As a demonstration, Fig.
\ref{fig:threshie} illustrates that an incident propagating plane
wave can couple into a state which is bound along the waveguide.
Note that the periodicity of the guided portion of the wave is not
the periodicity
of the lattice.%

\section{Arrays of Atoms as Waveguides\label{sec:Waveguides}}

We now apply the arguments of the previous sections to demonstrate
that a wall of atoms might be used to guide waves. By guiding, we
mean that if an incident beam is focused on some part of the
grating, the scattered wave propagates along the grating for a long
distance. In Section \ref{sub:Mechanisms-for-Guiding:} and Section
\ref{sub:Bound-States}, we discussed the existence of states that
are either quasibound or bound along the infinite array. These
states, as such, are not an example of transport; being
translationally invariant along the array, they can only be accessed
by initial conditions that are already translationally invariant,
such as plane waves. Transport along the array requires that the
array have a symmetry breaking point where an incident wave can be
injected. In particular, in order for a localized beam aimed at the
array to conduct, the array must have a defect or an end. The
asymmetry will of course affect the properties of the grating, e.g.
diffraction, impedance, etc. While we can use our results on the
infinite array as a basis for studies of guiding
in related systems, the infinite array itself is not a waveguide.%

\subsection{Guiding by a Semiinfinite Array\label{sub:Demonstration-of-Guiding}}

\begin{figure}

\begin{centering}\includegraphics[width=1\textwidth,keepaspectratio]{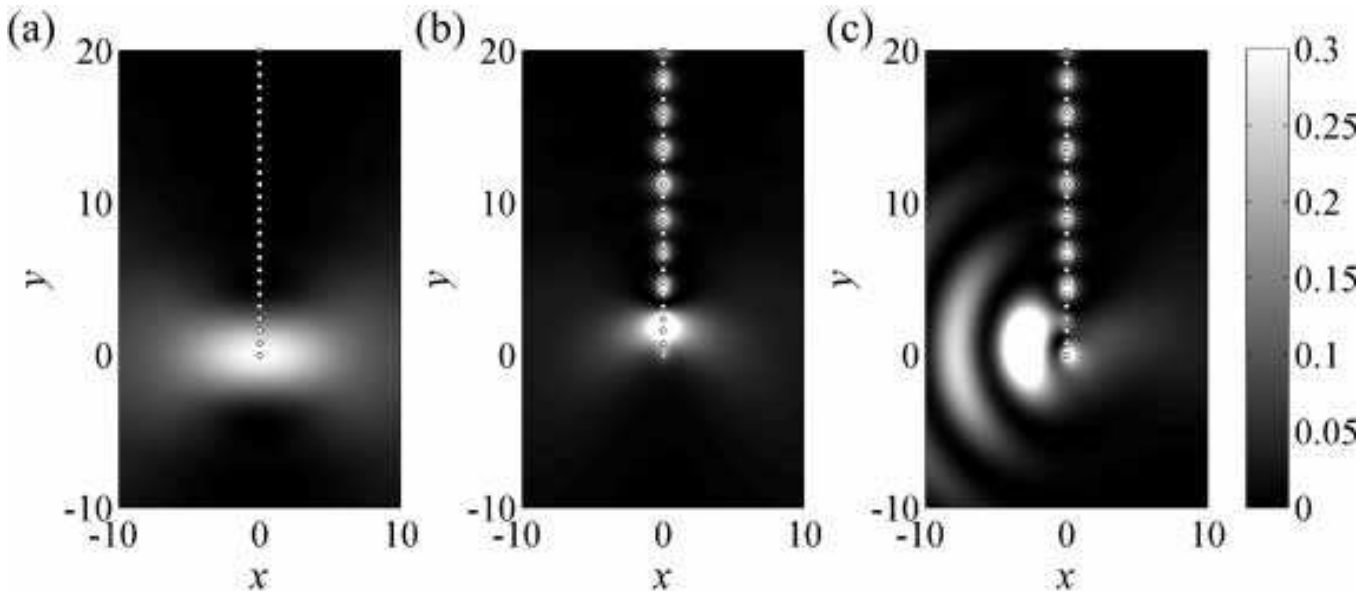}\par
\end{centering}

\caption{Guiding of a focused beam with $k=0.7$.
The scatterers are indicated by dots. These plots show the
probability densities of (a) the incident beam, focused on the end
of the array (b) The scattered wave, which is guided along the
array, and (c) the full scattered wavefunction (incident plus
scattered). The individual scatterers are attractive wells with the
same form and lattice constant as in Fig. \ref{fig:wallpole}(e)-(f).
\label{fig:focusguide}}

\end{figure}

Consider an incident beam \begin{equation}
\phi(\vec{r})=\frac{1}{2\pi}\int_{-\pi/2+\beta}^{\pi/2+\beta}g(\theta_{k}-\beta)e^{i\vec{k}(\theta_{k})\cdot\vec{r}}d\theta_{k},\label{eq:focused}\end{equation}
where $g(\theta_{k})=e^{-\left(\theta_{k}/w\right)^{2}}.$  Eq.
(\ref{eq:focused}) represents a beam incident from the right,
focused on the scatterer at $\vec{r}=0,$ and rotated by an angle
$\beta$ from the positive $x$ axis. An incident beam [Eq.
(\ref{eq:focused}) with $\beta=0$ (normal incidence)] is shown in
Fig. \ref{fig:focusguide}(a). As $w\rightarrow0$, $\phi(\vec{r})$
approaches a plane wave, and for $w\rightarrow\infty$,
$\phi(\vec{r})$ becomes a spherical wave with its form related to
the Fourier transform of the spherical wave $J_{0}(kr)$.  In the
following, we are interested in intermediate values of $w$, for
which Eq. (\ref{eq:focused}) represents a focused beam.

Using our results for the infinite array as a guide, we take a
purely numerical approach to the study of guiding in finite and
semiinfinite arrays, as these systems are difficult to treat
analytically. Analytical studies (e.g. Refs. \cite{Sommerfeld},
\cite{HankelSum}) typically conclude that scattering from a
semiinfinite array reproduces certain major features of an infinite
array: resonances, diffraction, etc. The main difference is that in
the semiinfinite case, a spherical wave emanates from the end of the
array. Most studies of semiinfinite arrays begin with the infinite
solution and derive this spherical edge wave
as a correction term.%

We demonstrate guiding numerically in Fig. \ref{fig:focusguide}. The
scatterers are closely spaced, and the focused beam is aimed at the
end of the array. In Fig. \ref{fig:focusguide}(c), a guided state
clearly propagates along the array. The guiding occurs at low
energies, near bound states of the infinite array. Guiding does not
occur for repulsive scatterers, confirming that the coupling is to
the true bound states of the array rather than the quasibound
states.

\begin{figure}

\begin{centering}\includegraphics[width=1\textwidth,keepaspectratio]{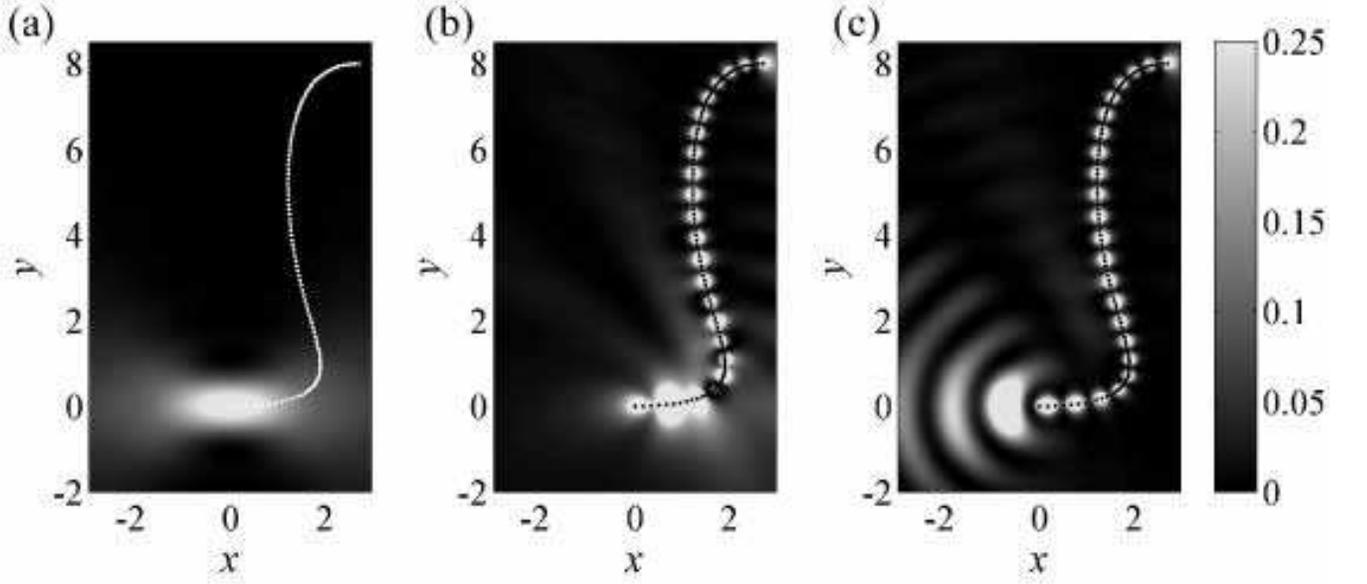}\par\end{centering}

\caption{Guiding of a focused beam along a finite
wall; $k=3\pi$. The scatterers (depicted by dots in the figure)
are attractive disks with $V_{0}=-10$, $a=0.2$. These plots show the
probability densities of (a) the incident beam, (b) the scattered
wave, and (c) the full scattered wavefunction (incident plus
scattered). In (b), amplitude visibly leaks away near the sharp bend
near $y=0$, as a consequence of nonadiabaticity.
\label{fig:curvedwall}}

\end{figure}

The guiding is robust and extends to gently curved and finite walls
as demonstrated in Fig. \ref{fig:curvedwall}, where a beam focused
on one end of a curved wall emerges at the other end. The scatterers
are effectively behaving like a light pipe for a matter wave. If the
wall is curved sharply relative to the wavelength of the incident
beam, amplitude leaks out as adiabaticity breaks down.

In this section, we have demonstrated via numerical simulations that
the $E>0$ conducting states of the array can be exploited to guide
waves along a semiinfinite wall of attractive, closely spaced scatterers.
The physical explanation is that a semiinfinite wall of attractive
potentials behaves like a trough, and furthermore that because the
trough has an end, it is possible to couple into conducting states
via injecting a beam into the end.

\subsection{Effective Cross Section for Conduction Along the Wire}

\begin{figure}
\begin{centering}\includegraphics[clip,width=0.3\textwidth,keepaspectratio]{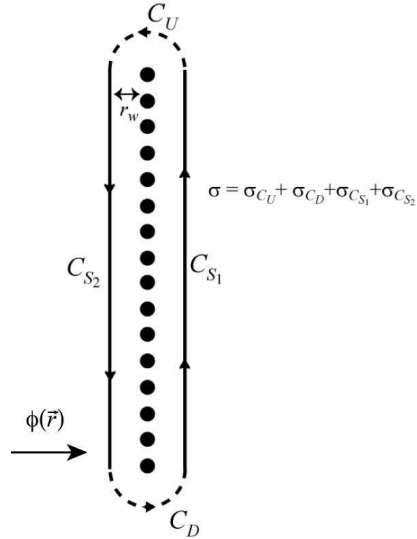} \par\end{centering}

\caption{The {}``stadium''-like contour used to evaluate the total
cross section, $\sigma$, for the wall of scatterers. The shortest
distance from any point on the contour to the wall is $r_{w}$. For
an incident beam focused on the bottom of the wall of scatterers
(indicated by the arrow), the fraction of particles
{}``transported'' along the wall is given roughly by
$\sigma_{C_{U}}/\sigma$. \label{fig:jamie1}}
\end{figure}

From Figs. \ref{fig:focusguide}-\ref{fig:curvedwall}, it appears an
incident beam of particles focused on one end of a wall of
scatterers can be efficiently guided to the other end. In this
section, we provide a quantitative estimate of this guiding
efficiency. Consider a focused beam $\phi(\vec{r})$ [Eq.
(\ref{eq:focused})] of energy $E=\frac{1}{2}k^{2}$ with normal
incidence ($\beta=0$) to the wall of scatterers.  Such an initial
incident beam has nonzero flux $j_{inc}$ along the
$x$ direction, where \[
j_{inc}=\int_{0}^{\pi}\text{d}\theta(k\cos(\theta))\exp\left(-2(\theta/w)^{2}\right).\]
The scattered flux can be calculated by evaluating the flux of the
scattered wave function $\psi_{s}(\vec{r})$ over any contour which
completely surrounds the scatterers. Finally, the total cross
section can be simply evaluated by integrating the scattered flux
over the desired contour and dividing by $j_{inc}$. For the wall of
scatterers, the total cross section, $\sigma$, can be calculated
over the contour shown in Fig. \ref{fig:jamie1}, giving
\begin{eqnarray}
\sigma & = & \frac{1}{j_{inc}}\int_{C}\text{d}\vec{a}\cdot\text{Im}[\psi_{s}^{*}(\vec{a})\vec{\nabla}\psi_{s}(\vec{a})]\nonumber \\
 & = & \sigma_{C_{U}}+\sigma_{C_{D}}+\sigma_{C_{S_{1}}}+\sigma_{C_{S_{2}}}\end{eqnarray}
 where $\sigma_{C_{U(D)}}$ is the contribution to the total cross
section from particles scattered through the upper (lower) semicircles.

As long the contour encloses all the scatterers, the value of
$\sigma$ is independent of the choice of contour. However,
$\sigma_{C_{U(D)}}$ and $\sigma_{C_{S_{1(2)}}}$ will depend on the
value of $r_{w}$ for the contour shown in Fig. \ref{fig:jamie1}. We
are interested in estimating the fraction of scattered particles
transported along the wire for an incident beam which is focused on
the bottom of the wall of scatterers, i.e., we are interested in the
ratio of $\sigma_{C_{U}}/\sigma$ (assuming each incident particle is
scattered). We should therefore choose $r_{w}$ such that
$r_{w}/L\ll1$, where $L=Nd$ is the length of the wall of $N$
scatterers with spacing $d.$

\begin{figure}
\begin{centering}\includegraphics[clip,width=0.5\textwidth,keepaspectratio]{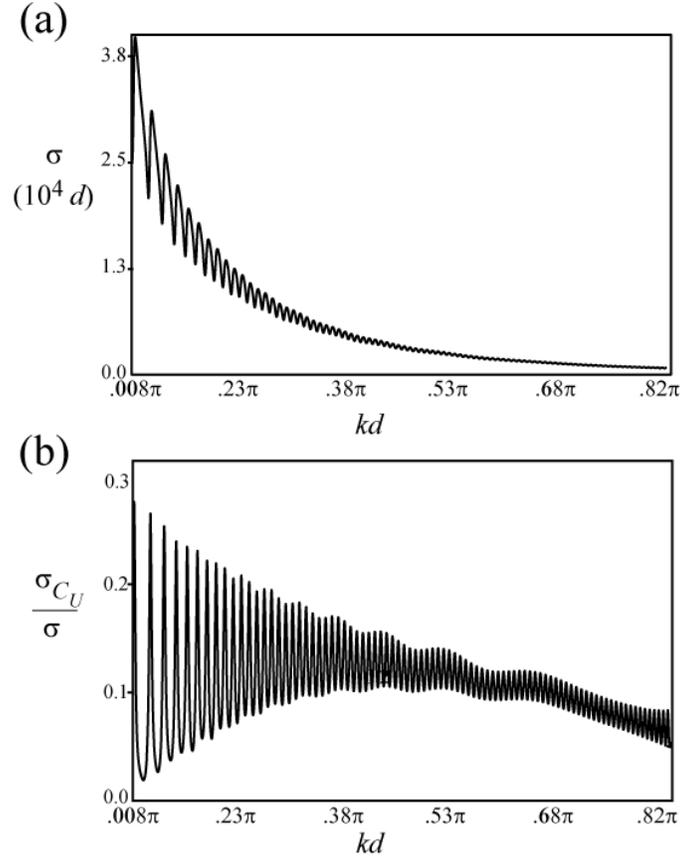} \par\end{centering}

\caption{(a) The total cross section, $\sigma$, in units of
$10^{4}d$ (where $d$ is the lattice spacing) for $N=150$ square well
scatterers ($V_{0}=-8$, $a=0.2$, $d=0.8$) as a function of $kd$. The
incident beam is focused upon the bottom scatterer, with $w=1$.
Since $\sigma$ is much greater than the spread of the incident beam,
it can be safely assumed that each incident particles are scattered.
Therefore in order to obtain the number of particles which are
transported along the scatterer wall, one must simply calculate
$\sigma_{C_{U}}/\sigma$, shown in (b). As can be seen, up to $30\%$
of all scattered particles can be transported along the wire
(compared with only $0.8\%$ if the scattering was assumed to be
isotropic).\label{fig:jamie2}}
\end{figure}

Figure \ref{fig:jamie2}(a) gives $\sigma$ (in units of $10^{4}d$) as
a function of $kd=\sqrt{2E}$ for a system of $N=150$ square well
scatterers characterized by $V_{0}=-8$, $d=0.8$, and $a=0.2$ (in
addition, $w=1$ was chosen for $\phi(\vec{r})$ in Eq.
(\ref{eq:focused})).   As can be seen from Fig. \ref{fig:jamie2}(a),
$\sigma$ increases with decreasing $kd$, although for the range of
$kd$ plotted in Fig. \ref{fig:jamie2}(a), $\sigma$ is still much
larger than the spatial spread of the incident wave function so that
essentially all of the incident particles are scattered. In order to
estimate the amount of transport along the scatterer wall, the ratio
of the scattered flux through the other end of the wall is taken
with the total scattered flux, which is simply equal to
$\sigma_{C_{U}}/\sigma$; this ratio is shown in Figure
\ref{fig:jamie2}(b). Here, the fraction of scattered particles which
are transported along the wire can reach up to 30\%, and oscillates
with $k$.  Note that the oscillations are related to the actual
length of the wall;  if the number of scatterers (and hence the
array's length) is doubled, the period of the oscillations would
roughly be doubled too.  For $0.38\pi\leq k \leq 0.53\pi$,
$\sigma_{C_{U}}/\sigma$ oscillates slightly about
$\sigma_{C_{U}}/\sigma\approx 12\%$.  This means that even for a
beam of particles with a distribution of incident energies (possibly
due to thermal effects), at least 12\% of the incident particles
will still be transported along the wire.  Also note that in these
calculations, $r_{w}=10d$ so that $r_{w}/L=1/40\ll1$. It is
interesting to observe that for a beam incident only on the
lowermost scatterer, and for the case of isotropic scattering (i.e.,
neglecting multiple scattering), the amount of scattered particles
transported through the same region relative to the total scattered
flux would be given by $2r_{0}/(2\pi L)\approx0.8\%$. Thus the wall
of scatterers can enhance the scattering along a given direction by
almost a factor of 40.

\section{Conclusions and Future Directions}

In this paper, we have examined scattering from and guiding by
quasi-1D periodic gratings of scatterers embedded in 2D. This system
is a good laboratory for highlighting the coexistence of scattering
phenomena, such as transmission, reflection, and resonance, with
features typical of periodic systems, such as band structure,
diffraction and conduction along the array.  Arranging individual
atoms on substrates, often in patterns far more intricate than
gratings, has become an established experimental technique
\cite{QuantumCorralsEigler,goldchains1,goldchains2,goldchains3}. The
motivation for our study is thus that atomic arrays can be built,
have been successfully modeled by multiple scattering theories,
\cite{GregCorrals,StellaThesis,QuantumCorralsRick}, and can
potentially serve as a waveguide for other particles.

In order to investigate the use of atomic arrays for guiding, we
have developed a multiple scattering theory (related to the KKR
method) for quasi-1D gratings of $s$ wave scatterers, embedded in
2D. The central physics is related to Bloch's theorem; we can obtain
the full scattered wavefunction from the solution for a single unit
cell by replacing the $t$ matrix of an individual scatterer with its
renormalized version and the free space Green's function with an
effective Green's function. We have used this result to discuss and
to examine some general features of scattering from a single chain
of atoms, such as resonances, quasibound states, and conducting states.
Finally, we have demonstrated numerically that conducting states of the
semiinfinite or finite array can be used to guide waves along
straight or curved walls with up to $30\%$ guiding efficiency.

The properties of the scattering and guiding studied here have
strong analogs in optical systems, such as planar waveguides,
optical fibers, and glass tabletops.  For the latter, light incident
on a sheet of glass at an angle and far from any edge is partially
reflected and transmitted directly. Some of the light is transmitted
and reflected indirectly, propagating inside the waveguide with
attenuation by incomplete internal reflection.  On the other hand,
light incident on a (symmetry breaking) edge can be partially
reflected, but mostly enters the sheet of glass and is trapped
inside by total internal reflection. Due to its resonant modes, our system has shown analogs to
each of these optical phenomena, and others.

This work can be extended in a number of directions. We have chosen
to embed our system in 2D because 2D is the relevant dimensionality
for electrons in surface states scattering from atoms adsorbed on
a metallic surface. With a change in the free space Green's function,
one could, however, revise the entire theory to treat a quasi-1D array
embedded in 3D, such as might be relevant if the scatterers were atoms
confined to an optical lattice. The physics of renormalization and
interference in a 3D system would be very similar, and presumably
identical guiding effects would arise.

Among the reasons that we have taken a multiple scattering approach
to the system is that it can be generalized to the introduction of
impurities or defects into the array. We have discussed how symmetry
breaking is required for guiding to occur. Introducing an impurity
or defect into an infinite array is one way of breaking symmetry,
with the impurity behaving as an antenna capable of drawing the
incident wave into the conducting state. The problem of impurities
in the array is also interesting in the sense that the array with an
impurity becomes a scattering system in the conducting mode,
yielding, in essence, a scattering theory within a scattering
theory. The question of guiding in a system with disorder may also
be of interest from the perspective of Anderson localization.
Interesting questions arise of how to maximize the guiding
efficiency, for example by impedance matching the array's end to
free space. Finally, Foldy's method has recently been extended to
include Rashba spin-orbit coupling \cite{JamieSO}, and it may be
possible to modify the results of this paper to examine spintronic
versions of guiding.

\section{Acknowledgments}

EJH and MA acknowledge support from the National Science Foundation
under NSF Grant No. CHE-0073544.

\bibliographystyle{apsrev}
\bibliography{atomwalls}

\appendix

\section{Relevant\label{sec:Relevant-Lattice-Sums} Lattice Sums}

Equation (\ref{eq:renormalized}) and Equation
(\ref{eq:latticehankelsum}) are in the form of spherical waves. We
would like to find a more useful expression for Eq.
(\ref{eq:latticehankelsum}), and also a more rapidly convergent
expression for Eq. (\ref{eq:renormalized}). Sums involving Hankel
functions converge to the plane wave limit very slowly. The physical
implication of this slow convergence is that edge effects are more
important than one might expect; many scatterers are required to
build an {}``infinite'' wall.

\subsection{Plane wave form of effective Green's function}

The Green's function for an infinite array of scatterers, as
expressed in spherical waves, is\begin{equation}
G(\vec{r})=\sum_{n=-\infty}^{\infty}G_{0}(\vec{r},\vec{r}_{n})e^{ik_{y}nd}.\label{eq:wallsphericalgreens}\end{equation}
Substituting the two dimensional free space Green's function [Eq.
(\ref{eq:freegreens})], we find that\begin{eqnarray}
G(\vec{r}) & = & -\frac{i}{2}\sum_{n=-\infty}^{\infty}H^{(1)}_{0}(k\left|\vec{r}-\vec{r}_{n}\right|)e^{ik_{y}nd}\label{eq:hankelgreens}\\
 & = & -\frac{i}{2\pi}\int_{-\infty}^{\infty}\frac{dk_{y}^{\prime}}{k_{x}^{\prime}}e^{i(k_{x}^{\prime},k_{y}^{\prime})\cdot(x,y)}\times\sum_{n=-\infty}^{\infty}\left[e^{-i(k_{y}^{\prime}-k_{y})d}\right]^{n}\nonumber \end{eqnarray}
where we have used the integral form of the Hankel function in the
right half plane $(x>0)$. Using\[
\sum_{n=-\infty}^{\infty}\left[e^{i(k_{y}^{\prime}-k_{y})d}\right]^{n}=\frac{2\pi}{d}\sum_{n=-\infty}^{\infty}\delta\left(k_{y}^{\prime}-k_{y}+\frac{2n\pi}{d}\right),\]
we can do the integral, yielding the sum of plane waves at real and
imaginary Bragg angles,\begin{eqnarray}
G(\vec{r}) & = & -\frac{i}{d}\sum_{n=-\infty}^{\infty}\frac{1}{k_{x}^{(n)}}e^{ik_{x}^{(n)}\left|x\right|}e^{ik_{y}^{(n)}y}\nonumber \\
 & = & -\frac{i}{d}e^{ik_{y}y}\sum_{n=-\infty}^{\infty}\frac{1}{k_{x}^{(n)}}e^{ik_{x}^{(n)}\left|x\right|}e^{\frac{-2in\pi}{d}y}\label{eq:planegreens}\end{eqnarray}
where we have taken the absolute value to ensure convergence, and\begin{eqnarray*}
k_{y}^{(n)} & \equiv & k_{y}-\frac{2n\pi}{d}\\
k_{x}^{(n)} & = &
\sqrt{k^{2}-\left(k_{y}^{(n)}\right)^{2}}\end{eqnarray*} define
wavevectors oriented at the Bragg angles. Imaginary values of
$k_{x}^{(n)}$ correspond to evanescent modes and are present only in
the near field.

This idea connects to the phenomenon of a {}``healing length'' when
examining reflections from a corrugated wall. We note that
$G(\vec{r})$ is singular as we approach the origin; while the
singularity is not evident in Eq. (\ref{eq:planegreens}), it is
readily apparent in Eq. (\ref{eq:hankelgreens}). Re-indexing, we
find\begin{equation}
G(\vec{r})=-\frac{i}{k_{x}d}e^{ik_{x}\left|x\right|}e^{ik_{y}y}-\frac{i}{d}e^{ik_{y}y}\sum_{n=1}^{\infty}\left(\frac{1}{k_{x}^{(-n)}}e^{ik_{x}^{(-n)}\left|x\right|}e^{\frac{2in\pi}{d}y}+\frac{1}{k_{x}^{(n)}}e^{ik_{x}^{(n)}\left|x\right|}e^{-\frac{2in\pi}{d}y}\right).\label{eq:reindexed}\end{equation}
The wavefunction is a Bloch wave:\[
\psi(\vec{r}+d\hat{y})=e^{ik_{y}d}\psi(\vec{r})\] with
$E=\frac{1}{2}k^{2}=\frac{1}{2}\sqrt{\left(k^{(n)}_{x}\right)^{2}+\left(k^{(n)}_{y}\right)^{2}}.$

\subsection{Kummer's Method: Extracting the Singularity}

As in Refs. \cite{MyPaper} and \cite{LatticeSums}, we want to apply
Kummer's method to extract the singularity from Eq.
(\ref{eq:planegreens}) and also to obtain a rapidly convergent
expression for the renormalized scattering strength, $\tilde{s}$
[Eq. (\ref{eq:renormalized})].  We begin rewriting Eq.
(\ref{eq:renormalized}) as
\begin{eqnarray}
G_{r} & = & \lim_{\vec{r}\rightarrow\vec{0}}\left[G(\vec{r})-G_{0}(\vec{r})\right]\nonumber \\
 & = & \lim_{\vec{r}\rightarrow\vec{0}}\left[\left(G(\vec{r})-S(\vec{r})\right)+\left(S(\vec{r})-G_{0}(\vec{r})\right)\right]\label{eq:kummerconverge}\end{eqnarray}
where $S(\vec{r})$ is a sum chosen to cancel the log singularity of
the Hankel function. Looking at Eq. (\ref{eq:planegreens}), we
choose the following form for $S(\vec{r}):$\[
S(\vec{r})=-\frac{1}{\pi}\sum_{n=1}^{\infty}\frac{1}{n}\times
e^{\frac{2n\pi}{d}\left|x\right|}\cos\left(\frac{2n\pi
y}{d}\right)\] With this choice of $S(\vec{r})$, we are extracting
from the zero energy limit from the singular sum in Eq.
(\ref{eq:reindexed}), evaluated (for simplicity) for a normally
incident. The choice of $S(\vec{r})$ is not unique;  our motivation
for choosing this particular form of $S(\vec{r})$ is that we know
that it will contain the logarithmic singularity.  From Eq.
(\ref{eq:wallsphericalgreens}) we see that the zero energy limit of
the Green's function corresponds to the limit of a Hankel function
as we approach the origin, and thus the zero energy limit of the sum
will be singular.

We shall proceed to show that although both $S(\vec{r})$ and $G(\vec{r})$
separately diverge logarithmically at the origin,\[
\lim_{\vec{r}\rightarrow\vec{0}}\left[S(\vec{r})-G_{0}(\vec{r})\right]\]
 is finite. We first find a closed form of $S(\vec{r})$:\begin{eqnarray*}
S(\vec{r}) & = & -\frac{1}{\pi}{\rm Re}\sum_{n=1}^{\infty}\frac{1}{n}\left(e^{\frac{2\pi}{d}\left|x\right|}e^{\frac{2i\pi y}{d}}\right)^{n}\end{eqnarray*}
We can now use \[
{\rm Re}\sum_{n=1}^{\infty}\frac{Z^{n}}{n}=-\frac{1}{2}\ln\left|1-Z\right|^{2}\]
to rewrite\begin{eqnarray*}
S(\vec{r}) & = & \frac{1}{2\pi}\ln\left[e^{\frac{2\pi}{d}\left|x\right|}\left(e^{\frac{-2\pi}{d}\left|x\right|}-e^{\frac{\pi}{d}\left|x\right|}e^{\frac{2i\pi y}{d}}\right)\left(e^{\frac{-2\pi}{d}\left|x\right|}-e^{\frac{\pi}{d}\left|x\right|}e^{\frac{-2i\pi y}{d}}\right)\right]\\
 & = & \frac{1}{2\pi}\ln\left\{ 2e^{\frac{2\pi}{d}\left|x\right|}\left[\cosh\left(\frac{2\pi}{d}x\right)-\cos\left(\frac{2\pi}{d}y\right)\right]\right\} \end{eqnarray*}
Using this form, it is simple to take the limit,\begin{equation}
\lim_{\vec{r}\rightarrow\vec{0}}S(\vec{r})=-\frac{1}{\pi}\ln\left(\frac{kd}{2\pi}\right)+\frac{1}{\pi}\ln
kr.\label{eq:slimit}\end{equation} Subtracting the limiting form of
the Hankel function\[
\lim_{\vec{r}\rightarrow\vec{0}}G_{0}(kr)=-\frac{i}{2}+\frac{\gamma-\ln2}{\pi}+\frac{1}{\pi}\ln
kr\] from Eq. (\ref{eq:slimit}), we find\[
\lim_{\vec{r}\rightarrow\vec{0}}\left[S(\vec{r})-G_{0}(\vec{r})\right]=-\frac{1}{\pi}\ln\left(\frac{kd}{4\pi}\right)+\frac{i}{2}-\frac{\gamma}{\pi}\]
 which is finite. Returning to Eq. (\ref{eq:kummerconverge}), we find
that another form of $G(\vec{r})$ is

\begin{eqnarray*}
G(\vec{r}) & = & -\frac{i}{d}\frac{1}{k_{x}}e^{ik_{x}\left|x\right|}-\frac{i}{d}\sum_{n=1}^{\infty}\left(\frac{1}{k_{x}^{(-n)}}e^{ik_{x}^{(-n)}\left|x\right|}e^{i\left(k_{y}+\frac{2in\pi}{d}\right)y}+\frac{1}{k_{x}^{(n)}}e^{ik_{x}^{(n)}\left|x\right|}e^{i\left(k_{y}-\frac{2in\pi}{d}\right)y}-\frac{d}{in\pi}e^{\frac{2n\pi}{d}\left|x\right|}\cos\left(\frac{2n\pi y}{d}\right)\right)\\
 &  & -\frac{1}{\pi}\ln\left(\frac{kd}{4\pi}\right)+\frac{i}{2}-\frac{\gamma}{\pi},\end{eqnarray*}
and from this expression, we can calculate $G_{r}$:\begin{eqnarray}
G_{r} & = & \lim_{r\rightarrow r_{0}}\left[G(\vec{r})-G_{0}(\vec{r})\right]\nonumber \\
 & = & \frac{-i}{k_{x}d}-\frac{i}{d}\mathop{\sum}_{n\neq0}\left(\frac{1}{k_{x}^{(n)}}-\frac{d}{2i\left|n\right|\pi}\right)-\frac{1}{\pi}\ln\left(\frac{kd}{4\pi}\right)+\frac{i}{2}-\frac{\gamma}{\pi}.\label{eq:grwall}\end{eqnarray}
 In the special case where the incident plane wave is normal to the
array, this expression is identical to $G_{r}$ for a scatterer at
any location in a periodic wire. In the general case, the value of
$G_{r}$ differs in the values of $k_{x}^{(n)},$ which now depend
on the incoming wavefunction.

\section{\label{sec:Unitarity-and-The}Unitarity and The Single Wall}

Simplifying Eqs. (\ref{eq:rn})-(\ref{eq:tn}) to the case of a single
chain of atoms, the reflection and transmission probabilities
are\begin{eqnarray}
R_{q} & = & -\frac{i\tilde{s}}{d}\frac{1}{\sqrt{k_{x}^{(0)}k_{x}^{(q)}}}\label{eq:wallrandt}\\
T_{q} & = & \delta_{q0}+R_{q}.\nonumber \end{eqnarray} The unitarity
requirement, $T+R=1$, can be shown to yield the following constraint
on $\tilde{s}$, which resembles the ordinary optical theorem, Eq.
(\ref{eq:freeopticalfoldy}):\begin{equation}
-\mbox{Im}\tilde{s}=\left|\tilde{s}\right|^{2}\sum_{q\in\mathcal{Q}}\frac{1}{k_{x}^{(q)}d}.\label{eq:walloptical}\end{equation}
Using Eq. (\ref{eq:stwiddle}) and Eq. (\ref{eq:grwall}), it is
straightforward to verify that $\tilde{s}$ satisfies the unitarity
condition, Eq. (\ref{eq:walloptical}), as long as the
single-scatterer $t$ matrix $s(k)$ satisfies the free space
unitarity condition, Eq. (\ref{eq:freeopticalfoldy}).
 In the limit
$kd\rightarrow0,$ our results approach those for a continuous wall
of scatterers. An alternate method for treating a continuous wall is
the boundary wall method \cite{boundarywall} where the wall is
discretized into a set of pseudo-scatterers.  The boundary wall
method has been shown to be equivalent to building a wall out of
point scatterers--with the exception that the self-interaction,
which was omitted in Eq. (\ref{eq:lippmanSchwingerMultiple2}), has
been left in the calculation, rendering the individual
pseudo-scatterer unphysical since its $t$ matrix no longer satisfies
the optical theorem in Eq. (\ref{eq:freeopticalfoldy}).  Our
formalism, though significantly more complicated, has the advantage
of corresponding physically to buiding a wall out of individual
atoms.

\section{Simulating Potentials With a t Matrix}
\subsection{Hard Disk}

The 2D hard disk is simulated by requiring that the wavefunction go
to zero at a radius $a$ from the scatterer. It is shown (see e.g.
\cite{HerschThesis}) that this corresponds to a $t$ matrix\[
s(k)=-2i\frac{J_{0}(ka)}{H_{0}^{(1)}(ka)}.\]

\subsection{Soft Disk}

Another example of a potential for which we can analytically calculate
the $s$ wave scattering properties is a soft disk of radius $a$,\begin{eqnarray}
V(r) & = & V_{0}\Theta(R-a).\label{eq:softdisk}\end{eqnarray}
We are particularly interested in the case $V_{0}<0,$ corresponding
to an attractive disk. %

The $s$ wave phase shift $\delta_{0}$ for the soft disk potential
is determined by continuity of the logarithmic derivative of the wavefunction
at $r=a:$\begin{equation}
\tan\delta_{0}=\frac{qJ_{0}(ka)J_{1}(qa)-kJ_{1}(ka)J_{0}(qa)}{qY_{0}(ka)J_{1}(qa)-kY_{1}(ka)J_{0}(qa)}\label{eq:phaseshift}\end{equation}
where\[
q\equiv\sqrt{2(E-V_{0})}\]
is the wavenumber inside the disk.

Having calculated the $s$ wave phase shift, it is straightforward to
confirm that we can simulate the $s$ wave scattering from a soft
attractive disk with a $t$ matrix\begin{equation}
s(k)=\frac{-2\tan\delta_{0}}{1-i\tan\delta_{0}}.\label{eq:stangent}\end{equation}
The negative energy poles of the $t$ matrix in Eq.
(\ref{eq:stangent}) occur at\begin{equation} qK_{0}(\kappa
a)J_{1}(qa)=\kappa K_{1}(\kappa
a)J_{0}(qa)\label{eq:bound}\end{equation} where
$\kappa\equiv\sqrt{-2E}.$ Eq. (\ref{eq:bound}) can easily be shown
to be the characteristic equation for bound states of the 2D
cylindrical well.
\end{document}